\documentclass[aps,prb,reprint,groupedaddress]{revtex4-1}

\usepackage{amsfonts,amssymb,amsmath,amsthm,graphicx}%,ulem}
\usepackage[dvipsnames]{xcolor}
\usepackage[colorlinks=true,linkcolor=red,urlcolor=cyan,citecolor=green]{hyperref}

\newcommand{\pd}[2]{\frac{\partial #1}{\partial #2}}

\newcommand{\csum}[0]{\sideset{}{'}\sum}

\begin{document}

%\preprint{}

\title{Local Volume Effects in the Generalized Pseudopotential Theory}


\author{Guy C. G. Skinner}
%\email[Electronic Address: ]{guy.c.skinner@kcl.ac.uk}
%\thanks{}
\affiliation{Department of Physics, King's College London, Strand,
  London WC2R 2LS, UK}

\author{John A. Moriarty}
%\email[Electronic Address: ]{guy.c.skinner@kcl.ac.uk}
%\thanks{}
\affiliation{Lawrence Livermore National Laboratory, Livermore,
  California 94551-0808}

%\author{Eleftherios I. Andritsos}
%\email[Electronic Address: ]{guy.c.skinner@kcl.ac.uk}
%\thanks{}
%\affiliation{Department of Physics, King's College London, Strand, London WC2R 2LS, UK}

\author{Anthony T. Paxton}
\email[Electronic Address: ]{tony.paxton@kcl.ac.uk}
%\thanks{}
\affiliation{Department of Physics, King's College London, Strand, London WC2R 2LS, UK}


\date{\today}

\begin{abstract}
  The generalized pseudopotential theory (GPT) is a powerful method for
  deriving real-space transferable interatomic potentials. Using a
  coarse-grained electronic structure, one can explicitly calculate
  the pair ion-ion and multi-ion interactions in simple and transition
  metals. Whilst successful in determining bulk properties, in central
  force metals the GPT fails to describe crystal defects for which there
  is a significant local volume change. A previous paper
  [\href{https://doi.org/10.1103/PhysRevLett.66.3036}{PhysRevLett.66.3036
      (1991)}] found that by allowing the GPT total energy to depend
  upon some spatially-averaged local electron density, the energetics
  of vacancies and surfaces could be calculated within experimental
  ranges. In this paper, we develop the formalism further by
  explicitly calculating the forces and stress tensor associated with
  this total energy. We call this scheme the adaptive GPT (aGPT) and
  it is capable of both molecular dynamics and molecular statics. We
  apply the aGPT to vacancy formation, divacancy binding and stacking
  faults in hcp Mg. We also calculate the local electron density
  corrections to the bulk elastic constants and phonon dispersion for
  which there is refinement over the baseline GPT treatment.  
\end{abstract}

% insert suggested keywords - APS authors don't need to do this %\keywords{}

\maketitle

\section{Introduction}
\label{sec:intro}

Generalized pseudopotential theory (GPT) is a first-principles
framework for deriving real-space interatomic potentials in metals and
alloys from density-functional quantum mechanics
\cite{Moriarty1977,*Moriarty1982,*Moriarty1988,Moriarty2019}. In a
basic plane-wave basis, the GPT provides an updated and refined
version of second-order pseudopotential perturbation theory, with
linear screening and nonlocal, energy-dependent pseudopotentials, that
can be applied to $sp$-valent, nearly-free-electron (NFE) simple
metals. More generally, in a mixed basis of plane waves and localized
atomic $d$ states, the GPT additionally captures both tight-binding
(TB) $d$-state interactions and $sp$-$d$ hybridization between the
broad NFE $sp$-bands and the narrow TB $d$ bands. The practical
challenges of the GPT for pure transition metals have also led to the
development of a simplified model GPT (MGPT)\cite{Moriarty1990} which
allows for large-scale atomistic simulations in these materials. The
GPT and MGPT have been successfully applied to transition-series
metals with empty, filled, and partially filled $d$-bands
\cite{Moriarty1977, Moriarty2019, Moriarty1990, Moriarty2006}, to
transition-metal alloys \cite{Moriarty2019, Moriarty1997}, and, with
localized $f$-states in place of the $d$-states, to actinide metals as
well \cite{Moriarty2019, Moriarty2006}.\par

For bulk elemental metals, the GPT total energy $E_{\rm tot}$ is
developed in a volume-dependent many-body cluster expansion
\cite{Moriarty1977, Moriarty2019}, which in its simplest form is
truncated at pairwise interactions: 

\begin{equation}
  E_{\rm tot} (\{ {\bf R} \}, \Omega) = N E_{\rm vol} (\Omega) +
  \frac{1}{2} \csum_{i,j} v_2 (R_{ij}, \Omega),
  \label{eq:etot}
\end{equation}

where $\Omega$ is the atomic volume and the prime on the double
summation over ion positions $i$ and $j$ excludes the $i = j$ term.
The large volume term $E_{\rm vol}$ is independent of the positions of
the ions, and accounts for most of the equilibrium cohesive energy of
the metal, as illustrated in Fig.\ref{fig:coh} for Mg. The functional
form of the smaller pair potential $v_2$ is also independent of atomic
structure, and $v_2 (R_{ij},\Omega)$ accounts for structural energy
differences between different configurations of the ions at volume
$\Omega$ through its explicit dependence on the ion-ion separation
distance $R_{ij} = | {\bf R}_j - {\bf R}_i|$. The GPT total-energy
functional given by Eq.\ref{eq:etot} well describes the bulk
properties of simple metals (e.g., Mg, Al), pre-transition metals with
nearby empty $d$ bands (e.g., Ca), late transition metals with nearly
filled $d$ bands (e.g., Ni, Cu), and post-transition metals with
completely filled $d$ bands (e.g., Zn). For the remaining central
transition metals, it is necessary to extend the total-energy
expansion in Eq.\ref{eq:etot} to include angular-dependent three- and
four-ion potentials, which are established, respectively, by the
third- and fourth-order moments of the $d$-band density of electronic
states. Computationally, the evaluation of the GPT total energy for
all metals scales linearly with the number of atoms and is thus an
order-$N$ process. For the non- and late-transition elements covered
by Eq.\ref{eq:etot}, however, there is an additional computational
overhead relative to short-ranged central-force empirical potentials
as a result of the long-ranged screening oscillations in the GPT pair
potential $v_2$.  Even so, this is not a significant barrier in most
applications today, and using modern high-performance computers,
large-scale GPT atomistic simulations involving millions of atoms can
be routinely performed \cite{Moriarty2019,Moriarty2006}. \par

The structure-independent nature of the pair and multi-ion potentials
in the GPT ensures that these potentials are transferable to all ion
configurations of the bulk metal, either ordered or disordered.  This
includes all structural phases of both the solid and the liquid, as
well as the deformed solid and imperfect bulk solid with either point
or extended defects present.  At the same time, the explicit volume
dependence of the volume term and potentials is \textit{global} and
not \textit{local}, so that the creation of a free surface, or even a
bulk defect that comes with significant free volume, such as a
vacancy, still receives no contribution to its formation energy from
$E_{\rm vol}$ in Eq.\ref{eq:etot}. As a result, both surface energies
and the vacancy formation energy can be significantly underestimated.
In simple metals, the problem with the vacancy formation energy in
particular is a well-known shortcoming of conventional second-order
pseudopotential perturbation theory \cite{Finnis2003}, as we further
discuss below in Sec. \ref{sec:proto} in the context of our present Mg
prototype. \par

To address such shortcomings in the GPT, Moriarty and Phillips
\cite{Moriarty1991} transformed the bulk global-volume representation
of the total energy to an equivalent local electron-density
representation, such that Eq.\ref{eq:etot} becomes

\begin{equation}
  E_{\rm tot} (\{ {\bf R} \}, n_{\rm val}) = \sum_i E_{\rm vol}
  (\bar{n}_i) + \frac{1}{2} \csum_{i,j} v_2 (R_{ij}, \bar{n}_{ij}),
  \label{eq:agpt}
\end{equation}

where $\bar{n}_i$ is a simple functional of the average value of the valence
electron density $n_{\rm val}$ on the site $i$, and $\bar{n}_{ij}$ is
the arithmetic average $(\bar{n}_i + \bar{n}_j)/2$. For central
transition metals there are corresponding three- and four-ion
potential contributions on the right-hand-side of Eq.\ref{eq:agpt}. In
the perfect crystal with equivalent ion positions, Eq.\ref{eq:agpt} is
an exact transformation and only a redefinition of variables, with all
quantities still determined from first principles. The step forward
comes in then, as an \textit{ansatz}, applying Eq.\ref{eq:agpt} to all
ion configurations, including free surfaces and bulk defects. In doing
so, one notes from Fig. 1 that qualitatively the missing positive
formation energy for surfaces and vacancies is indeed now supplied by
the volume term, because $\bar{n}_i$ is lower near a surface or
vacancy site than at a bulk ion site. Moriarty and Phillips went on
to show that good \textit{unrelaxed} surface energies and vacancy
formation energies could thereby be obtained for both the late
transition metal Cu and for the central transition metal Mo. In the
case of Cu, the local-density corrections were found to be very large,
averaging about 70\% for both the surface energies and for the vacancy
formation energy. In the case of Mo, on the other hand, the
corrections were found to be significantly smaller, 30-40\% for the
surface energies and only 5\% for the vacancy formation energy.  The
physical reason for the latter behavior is that in transition metals
the essential local character needed in the total energy is already
present to a large degree in the global-volume representation through
the $d$ bonding contributions to $E_{\rm tot}$ provided by the
localized $d$-state moments. Thus for central transition metals, one
expects that bulk defect energies will be well calculated by either the
global-volume or the local-density formulations of the total
energy. \par

In the present paper, we take an additional major step and develop the
local-density representation of GPT into a robust general method we
now call the \textit{adaptive} GPT or aGPT, which includes not only
energies but the forces and stresses needed for atomistic simulation
and a much wider treatment of materials properties. The formalism of
the aGPT is elaborated in Sec. \ref{sec:agpt}, including the averaging
required in Eq.\ref{eq:agpt}. For simplicity this discussion is done
in the context of a well-studied simple-metal Mg prototype (see
Sec. \ref{sec:proto}), but the results can be immediately applied to
the empty, almost filled, and filled $d$-band metals covered by
Eqs.\ref{eq:etot} and \ref{eq:agpt}. The averaging is not unique, but
it can be optimized, and we have developed a good way to do this that
makes calculated properties quite insensitive to the parameters
defining the averaging, while at the same time allowing the
calculation of smooth derivatives of the averaged quantities. In
Sec. \ref{sec:derivs} we discuss the evaluation of the corresponding
aGPT forces and stresses, and test the results with calculations of
phonons and elastic constants. In this regard, an earlier, simplified
form of Eq.\ref{eq:agpt} was used by Rosenfeld and Stott
\cite{Rosenfeld1987} to resolve the well-known bulk compressibility
problem in pseudopotential perturbation theory, as we further discuss
below in Sec. \ref{sec:bulk}, and as we use here as an additional
fundamental test for the aGPT elastic moduli. Finally, in
Sec. \ref{sec:results} we apply the aGPT to the calculation of relaxed
single vacancies and divacancies as well as to stacking fault energies
in hcp Mg.  

\subsection{The Bulk Compressibility Problem}
\label{sec:bulk}

There are two fundamental ways to calculate the bulk modulus of a
single crystal using the interatomic pair potentials derived from
second-order pseudopotential perturbation theory, or more generally
from the GPT. The first method involves taking the explicit second
volume derivative of the total energy given by Eq.\ref{eq:etot}. This
procedure corresponds to a homogeneous deformation of the primitive
cell of the lattice and produces the so-called static bulk modulus $B_s$.
The second method calculates the bulk modulus using the
long-wavelength (low-${\bf q}$) behavior of the dynamical matrix,
which determines the elastic constants of the material
\cite{Born1954}. This produces the so-called dynamic bulk modulus
$B_d$. These two methods are known to disagree over the value of the
bulk modulus produced. This discrepancy can be seen immediately to be
the result of the absence of explicit volume derivatives in the
dynamical matrix. In conventional pseudopotential perturbation theory,
the discrepancy is only resolved at fourth order
\cite{Brovman1970,*Brovman1974}, albeit in a computationally
challenging and non-transparent manner.  It was later shown
\cite{Rosenfeld1987}, by allowing the total energy to depend on local
electron density as in Eq.\ref{eq:agpt}, that the requisite volume
derivatives arise to correct the bulk modulus calculated from the
dynamical matrix. In the present context, one can use the accurate
value of $B_s$ calculated from Eq.\ref{eq:etot} to test the value of
$B_d$ calculated with the aGPT from Eq.\ref{eq:agpt}.

\subsection{Magnesium Prototype and Baseline Vacancy Formation Energy}
\label{sec:proto}

\begin{figure}[t]            
  \includegraphics[width=0.45\textwidth]{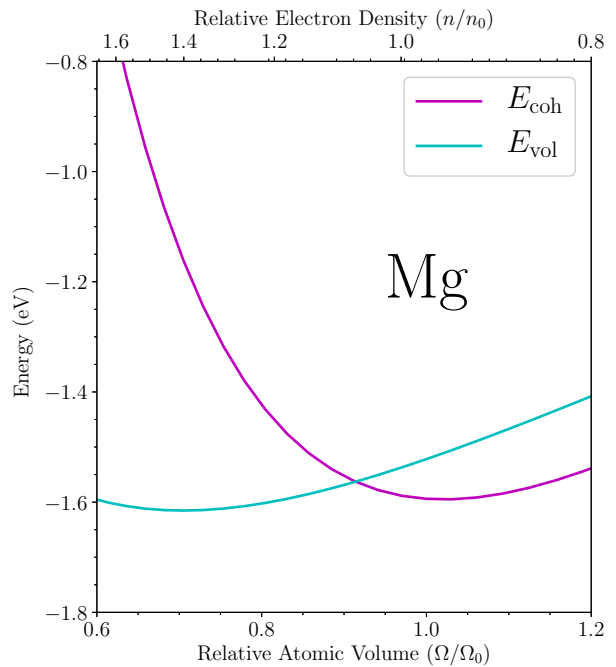}
  \caption{\label{fig:coh}GPT cohesion curve $E_{\rm coh} = E_{\rm
      tot} / N$ and volume term $E_{\rm vol}$ for Eqs.\ref{eq:etot}
    and \ref{eq:agpt}, as calculated from first principles for the
    simple metal Mg. Here $\Omega_0 = 156.8 ~\rm{a.u.}$ is the
    observed equilibrium volume and $n_0 = Z/\Omega_0$ is the
    corresponding average valence electron density for the bulk, with
    $Z = 2$.}
\end{figure}

Magnesium is an important lightweight metal whose bulk properties are
very well described by the GPT via Eq.\ref{eq:etot}, making it an
excellent prototype material for developing the aGPT.  The
first-principles pair potentials $v_2$ and volume term $E_{\rm vol}$
for this metal have been calculated over a wide volume range in
connection with detailed studies of the temperature-pressure phase
diagram and thermodynamic properties of Mg in the mid 1990s
\cite{Althoff1993,*Moriarty1995}, and in subsequent studies of
thermoelasticity \cite{Greeff1999}. The volume term is that displayed in
Fig.\ref{fig:coh}, and the Mg pair potentials used in this paper are
the same as in Refs. \cite{Althoff1993, *Moriarty1995} and
\cite{Greeff1999} except for an improved smooth long-ranged cutoff
function discussed in Sec. \ref{sec:agpt}.  As can be appreciated from
Fig.\ref{fig:coh}, good elementary cohesive properties are predicted,
including the cohesive energy, hcp lattice constant, and static bulk
modulus. The latter has a value $B_s = 35.8$ GPa at the observed
equilibrium volume in good agreement with the measured experimental
value of 35.2 GPa \cite{Slutsky1957}.  The calculated hcp phonon
spectrum is in excellent agreement with experiment, as are the
high-temperature values of the thermal expansion coefficient, specific
heat, and Gr\"{u}neisen parameter. Structural phase stability is well
predicted including the observed ambient pressure hcp structure with a
$c/a$ ratio near its observed value of 1.62, as well as the observed hcp
$\to$ bcc phase transition near 50 GPa.  Finally, the ambient pressure
melting properties are very well described, and the high-pressure melt
curve has been calculated to 50 GPa. \par

Also of interest in developing the aGPT is the baseline value of the
unrelaxed vacancy formation energy at constant volume, $\Omega =
\Omega_0$, as calculated from Eq.\ref{eq:etot} in the bulk GPT. This
quantity is given by \cite{Moriarty2019}

\begin{align}
  E^{\rm u}_{\rm vac} &= -\left(E_{\rm coh}^0 - E_{\rm vol}^0\right) +
  \Omega_0 P_{\rm vir}^0, \nonumber \\
  &= - \frac{1}{2} \sum_{i \ne 0} v_2 \left(R_i^0, \Omega_0\right) -
  \frac{1}{6} \sum_{i \ne 0} R_i^0 \pd{v_2 (R^0_i, \Omega_0)}{r},
  \label{eq:vac}
\end{align} 

where $E_{\rm coh}^0 \equiv E_{\rm tot} (\{{\bf R^0}\}, \Omega_0)/N$
and $E_{\rm vol}^0 \equiv E_{\rm vol} (\Omega_0)$.  The virial
pressure $P_{\rm vir}^0 \equiv P_{\rm vir} (\Omega_0)$  arises in
connection with the energy needed to compress the lattice uniformly
and maintain constant volume $\Omega = \Omega_0$  once the vacancy is
created.  Of the two terms on the second line of Eq.\ref{eq:vac}, the
second virial pressure term is the largest for Mg, but the total is
only $E_{\rm coh}^u = 0.44 ~{\rm eV}$, some 45\% below the measured vacancy
formation energy, as discussed in Sec. \ref{sec:results}.

\section{Formalism of the aGPT}
\label{sec:agpt}

\subsection{Treatment of the Electron Density in the GPT}
\label{sec:electroGPT}

The local volume change associated with a crystal defect gives rise to
a local change in the valence electron density. We briefly review the
treatment of the electron density in the GPT as applied to $sp$-valent
simple metals. The valence electron density consists of a uniform
electron density $n_{\rm unif} = Z / \Omega$ (where $Z$ is the
valence) plus small oscillatory and charge-neutral screening and
orthogonalization-hole components 
\cite{Moriarty1977,Moriarty1982,Moriarty1988,Moriarty2019}

\begin{equation}
  n_{\rm val} ({\bf r}) = n_{\rm unif} + \delta n_{\rm scr} ({\bf r})
  + \delta n_{\rm oh} ({\bf r}).
  \label{eq:nval}
\end{equation}

The screening electron density $\delta n_{\rm scr}$ arises from
first-order pseudopotential perturbation theory which for a simple
metal has the form\cite{Moriarty2019}

\begin{equation}
  \delta n_{\rm scr} ({\bf r}) = \csum_{\bf q} S ({\bf q}) n_{\rm scr}
  (q) e^{i {\bf q} \cdot {\bf r}}
  \label{eq:nscr}
\end{equation}

where $S ({\bf q}) = N^{-1} \sum_i \exp (-i {\bf q} \cdot {\bf R}_i)$
is the structure factor and\cite{Moriarty2019}  

\begin{equation}
  n_{\rm scr} (q) = -\left( \bar{w}_{\rm ion} (q) + \frac{4\pi
    e^2}{q^2} \left(1 - G (q)\right) n_{\rm oh} (q) \right)
  \frac{\Pi_0 (q)}{\epsilon (q)} 
  \label{eq:nscrq}
\end{equation}

where $\bar{w}_{\rm ion}$ is a well-defined average value of the ionic
pseudopotential over the free-electron Fermi sphere, $G$ is the
exchange-correlation functional, $\epsilon$ is the dielectric function
of the interacting electron gas and $\Pi_0$ is the electron gas
polarizability in the Hartree or random phase approximation. Each of
these quantities can be directly evaluated in terms of input
pseudopotential and electron gas quantities. \par

The orthogonalization-hole component arises from the difference
between the valence electron density constructed from the one-particle
pseudowavefunctions and the valence electron density constructed from
the `true' one-particle wavefunctions. For the non-local,
energy-dependent Austin-Heine-Sham (AHS)
pseudopotential\cite{Austin1962} used in the GPT, there exists an
exact transformation between the one-particle pseudo- and `true'
wavefunctions which can be exploited to obtain the exact
orthogonalization-hole density. The orthogonalization-hole
contribution to $n_{\rm val}$ in Eq.\ref{eq:nval} has the
form\cite{Moriarty2019}   

\begin{equation}
  \delta n_{\rm oh} ({\bf r}) = \left( \frac{Z^\ast}{Z}-1 \right)
  n_{\rm unif} + \sum_i n_{\rm oh} ({\bf r} - {\bf R}_i) 
  \label{eq:noh}
\end{equation}

where $Z^{\ast}$ is an effective valence occupation ($Z^\ast \geq
Z$) and $n_{\rm oh}$ is a localized hole density. For a simple metal,
$n_{\rm oh}$ is confined to the inner-core region of the site $i$, but
both $Z^\ast$ and $n_{\rm oh}$ depend on the properties of the
pseudopotential. For the non-local, energy-dependent AHS
pseudopotential used in the GPT, we have\cite{Moriarty2019} 

\begin{equation}
  Z^\ast = Z + \frac{2\Omega}{(2\pi)^3} \int {\rm d} {\bf k} ~\langle
  {\bf k} | p_c | {\bf k} \rangle \Theta^< (k - k_f) 
\end{equation}

and 

\begin{widetext}
  \begin{equation}
    n_{\rm oh} ({\bf r}) = \frac{2\Omega}{(2\pi)^3} \int {\rm d} {\bf
      k} ~\left[ \langle {\bf r} | p_c | {\bf k} \rangle \langle {\bf
        k} | p_c | {\bf r} \rangle - \langle {\bf r} | p_c | {\bf k}
      \rangle \langle {\bf k} | {\bf r} \rangle - {\rm c.c.} \right]
    \Theta^< (k_f - k) 
  \end{equation}
\end{widetext}

where $\Theta^<$ is a Heaviside step function that ensures that the
integral is over just the free-electron Fermi sphere and $p_c$ is the
inner-core projection operator  

\begin{equation}
  p_c = \sum_c | \phi_c \rangle \langle \phi_c |.
\end{equation}

The valence electron density $n_{\rm val}$ can equivalently be written
as a superposition of self-consistently screened pseudoatom densities
$n_{\rm pa}$ 

\begin{equation}
  n_{\rm val} ({\bf r}) = \sum_i n_{\rm pa} ({\bf r} - {\bf R}_i).
  \label{eq:nvpa}
\end{equation}

The precise form of the pseudoatom density $n_{\rm pa}$ can readily be
derived from Eqs.\ref{eq:nval}, \ref{eq:nscr} and \ref{eq:noh}. We do
this by inserting the full-form of the structure factor $S ({\bf q})$
into Eq.\ref{eq:nscr} and then adding the ${\bf q} = 0$ term to the
summation over ${\bf q}$ to account for the net uniform density
$Z^\ast n_{\rm unif}/Z$. Finally, we convert the summation over ${\bf
  q}$ to an integral and infer that the single-site pseudoatom density
is given by 

\begin{equation}
  n_{\rm pa} ({\bf r},\Omega) = \frac{\Omega}{(2\pi)^3} \int {\rm
    d}{\bf q} ~n_{\rm scr} (q) e^{i {\bf q} \cdot {\bf r}} + n_{\rm
    oh} ({\bf r}). 
  \label{eq:npa}
\end{equation}
  
\begin{figure}[t]            
  \includegraphics[width=0.45\textwidth]{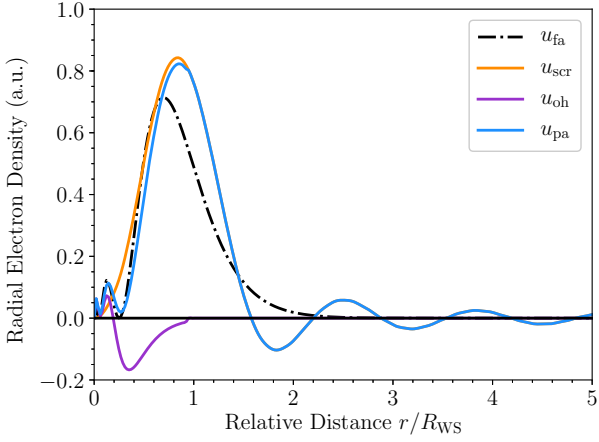}
  \caption{\label{fig:raddens}The radial valence electron density
    $u(r) = 4 \pi r^2 n (r)$ for the $3s$ and $3p$ bands in Mg for the
    pseudoatom $n = n_{\rm pa}$ (blue) in the bulk metal at $\Omega =
    156.8 ~{\rm a.u.}^3$ and also for the corresponding free atom $n =
    n_{\rm fa}$ (checked). The pseudoatom valence electron density
    replicates the inner-core density oscillations of the
    free-atom. At larger distances from the ion, the pseudoatom
    valence electron density is pushed outward relative to the
    free-atom, and has the familiar Friedel long-range screening
    oscillations. Also shown are the real-space screening density
    $n_{\rm scr}$ (orange) and orthogonalization-hole density (purple).} 
\end{figure}

The calculated GPT radial pseudoatom density $u_{\rm pa} ({\bf r}) = 4
\pi r^2 n_{\rm pa} ({\bf r})$ for Mg at the experimental
room-temperature atomic volume $\Omega = 156.8 ~\rm{a.u.}$ is shown in
Fig.\ref{fig:raddens} and compared with the corresponding free-atom
density for the valence $3s$ and $3p$ electrons. In this calculation,
and all those subsequent, the exchange-correlation functional $G (q)$
is taken to be the analytic expression developed by Ichimaru and
Utsumi \cite{Ichimaru1981} referenced to the exchange-correlation
energy of Vosko {\it et al.} \cite{Vosko1980} Whilst the discussion in
this section has been limited to $sp$-valent simple metals, the
extension to empty, filled and partially-filled $d$-band metals
covered by Eqs. \ref{eq:etot} and \ref{eq:agpt} does not alter the
subsequent discussion. \par 

\subsection{Implementing the aGPT}
\label{sec:imp}

To connect the GPT valence electron density with the aGPT total energy
in Eq.\ref{eq:agpt}, we spatially average the GPT valence electron
density $n_{\rm val}$ about the site $i$ using an arbitrary normalized
distribution function $f_w$. For a bulk crystal with equivalent ion
sites, the spatially-averaged electron density $\bar{n}_i$ about a
site $i$ is constrained to be the uniform valence electron density
$n_{\rm unif}$. Combining the two equivalent valence electron density
formulations in Eqs.\ref{eq:nval} and \ref{eq:nvpa}, yields the {\it
  bulk constraining equation}  

\begin{equation}
  \bar{n}_i \equiv n_{\rm unif} = \sum_j \bar{n}_{\rm pa}
  (R_{ij},\Omega) - \delta \bar{n}_{\rm oh}^i - \delta \bar{n}_{\rm
    scr}^i. 
  \label{eq:bulk}
\end{equation}

Here the bar over the densities refer to an averaging with respect to
some distribution function $f_w$ i.e.

\begin{equation}
  \bar{n}_{\rm pa} (R_{ij},\Omega) = \int {\rm d}{\bf r} ~f_w ({\bf r}
  - {\bf R}_i) n_{\rm pa} ({\bf r} - {\bf R}_j, \Omega) 
  \label{eq:ave}
\end{equation}

with both $\delta \bar{n}^i_{\rm scr}$ and $\delta \bar{n}^i_{\rm oh}$
having similar forms. Typically, this averaging smooths out the
long-range screening oscillations. As a result of the
bulk constraining equation, the aGPT preserves the bulk total energy
for any given crystal structure with equivalent ion sites. \par

The first step towards developing a practical aGPT scheme for
describing defects or surfaces is to make the approximation that

\begin{equation}
  \bar{n}_i = \sum_j \bar{n}_{\rm pa} (R_{ij},\Omega) - \delta
  \bar{n}_{\rm oh}^i - \delta \bar{n}_{\rm scr}^i 
\end{equation}

can be applied generally. Furthermore, the spatially-averaged local
electron density $\bar{n}_i$ can be broken down into an effective
on-site contribution $\bar{n}^i_a = \bar{n}_{\rm pa} (R_{ii},\Omega) -
\delta \bar{n}_{\rm oh}^i - \delta \bar{n}_{\rm scr}^i$ and an
off-site or background component $\bar{n}^i_b$ where 

\begin{equation}
  \bar{n}^i_b = \sum_{j \ne i} \bar{n}_{\rm pa} (R_{ij},\Omega).
\end{equation}

We make an additional assumption that the on-site density is constant
$\bar{n}^i_a \equiv \bar{n}_a$ and as a result only the background
density $\bar{n}_b^i$ is site-dependent. Under these assumptions, we
may calculate the on-site density $\bar{n}_a$ using the bulk
constraining equation in Eq.\ref{eq:bulk}. In practice this amounts to
first calculating $\bar{n}_a$ for an ideal bulk crystal prior to
calculating the total energy for the surface or defective
crystal. For certain $d$-band metals e.g. Cu, there may be $s$-$d$
transfer between the surface and the bulk \cite{Moriarty1991}. In
which case, all of the densities $\bar{n}_i$, $\bar{n}_a^i$ and
$\bar{n}_b^i$ must be scaled by a factor $Z_i / Z$ to account for
this, where $Z_i$ is an effective $sp$ occupation on the site
$i$. This quantity would have to be determined self-consistently. \par

The next step towards a practical aGPT implementation is to specify
the form of the distribution function $f_w$ in Eq.\ref{eq:ave}. We
choose $f_w$ to correspond to a sigmoid function

\begin{equation}
  f_w (r) = 
  \begin{cases}
    {\cal N}^{-1} & r < R_a
    \\
    {\cal N}^{-1} \left(1 + \alpha \left( \frac{r}{R_0} - 1
    \right)^2\right) e^{-\alpha \left( \frac{r}{R_0} - 1 \right)^2} &
    r \geq R_a
  \end{cases}
  \label{eq:sig}
\end{equation}

which is the sigmoid function that is typically used in the GPT to
truncate the pairwise interaction\cite{Yang2001} albeit with a
different value of the Gaussian width $\alpha$. For large values of
$\alpha$ this corresponds to an average over a sphere of radius
$R_a$. The normalization $\cal N$ of the distribution function $f_w$
is given by

\begin{equation}
  {\cal N} = V_w + \frac{8 \pi}{\alpha} R_a^{3} + \frac{5 \pi^{3/2}}{2
    \alpha^{3/2}} R_a^{3} + \frac{3 \pi^{3/2}}{\sqrt{\alpha}} R_a^3
\end{equation} 

which in the limit $\alpha \to \infty$ is the volume of a sphere of
radius $R_a$. The two parameters $\alpha$ and $R_a$ represent the only
parameters in this form of the aGPT. The Gaussian width $\alpha$ is
chosen such that the radial derivatives of the spatially averaged
pseudoatom density are smooth. If the radial derivatives were not
smooth then there would be an unphysically large change in the forces
as the interatomic separation changes from less than $R_a$ to greater
than $R_a$ and vice-versa. In the rest of this paper, we choose
$\alpha = 25$ which produces a spherically averaged pseudoatom density
with smooth derivatives over a wide range of averaging sphere
radii. We have a certain amount of freedom in choosing a value for
$R_a$ since physical properties of interest do not seem strongly
dependent on $R_a$. We choose the optimum of $R_a$ to be that which
reproduces the GPT volume-conserving elastic constants most
closely. These issues will be discussed further in Sections
\ref{sec:derivs} and \ref{sec:results}. Whilst other normalized
distribution functions have been trialled, none represented an
improvement on the sigmoid function. \par

\begin{figure}[t]            
  \includegraphics[width=0.45\textwidth]{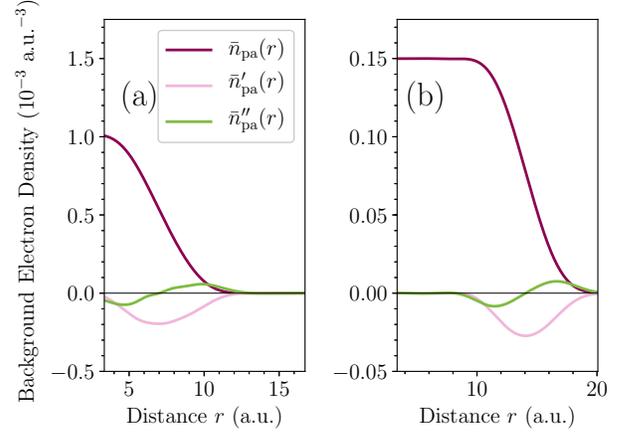}
  \caption{\label{fig:back}The spatially-averaged pseudoatom density
    $\bar{n}_{\rm pa}$ (magenta) is calculated for Mg at the
    experimental room-temperature atomic volume $\Omega = 156.8 ~{\rm
      a.u.}$. The radial derivatives (green and pink) of the
    spatially-averaged pseudoatom density were calculated using
    Lagrange interpolation polynomials. All of these quantities are
    calculated at $\alpha = 25$ and for two values of the averaging
    sphere radius $R_a = 1.8 R_{\rm WS}$ in (a) and $R_a = 3.4 R_{\rm
      WS}$ in (b).}
\end{figure}

The resulting spatially-averaged pseudoatom density is shown in
Fig.\ref{fig:back}. For values of $R_a$ in the range $R_a/R_{\rm WS}
\in [1, 2]$, where $R_{\rm WS} = (3 \Omega / 4 \pi)^{1/3}$ is the
Wigner-Seitz radius, the spatially-averaged pseudoatom density looks
like a Gaussian. A function of this type was proposed in the empirical
approach taken previously\cite{Rosenfeld1987,Finnis1998}. For larger
values of $R_a$ in the range $R_a/R_{\rm WS} \in [3, 4]$, the
resulting spatially-averaged pseudoatom density is almost flat over
the first two neighbor shells. \par

The spatially averaged pseudoatom density is smoothly truncated to
ensure force continuity during molecular dynamics. If we denote $R_0$
and $R_c$ as the cut-off onset and final termination respectively,
then our approach is to replace $\bar{n}_{\rm pa}$ by a polynomial
whose value and derivatives exactly match $\bar{n}_{\rm pa}$ at $R_0$
and whose derivatives are precisely zero at $R_c$. This polynomial can
be found using Hermite interpolation\cite{Burden2010} which finds an
$(n m -1)$ interpolating polynomial given knowledge of the function
and $m-1$ derivatives at $n$ points. For our purposes, we choose $m =
3$, $n = 2$ and $R_c - R_0 = 0.5 R_{\rm WS}$. 

\section{Total Energy Derivatives}
\label{sec:derivs}

\subsection{Forces \& Force Constants}
\label{sec:forces}

The force $F_{i\alpha}$ on the atom $i$ describes how the total energy
changes with respect to an infinitesimal shift in its position
$R_{i\alpha}$. As the ion-ion potential is self-consistently screened,
we can ignore any change in electron screening\cite{Feynman1939} 

\begin{equation}
  F_{i\alpha} = - \pd{E_{\rm tot}}{R_{i\alpha}} (\{{\bf R}\},\Omega).
\end{equation}

The force in the GPT involves only radial derivatives of the screened
ion-ion interaction whereas the aGPT force will involve contributions
from density derivatives of both $E_{\rm vol}$ and $v_2$. It is
instructive to decompose the force into three parts 

\begin{equation}
  F_{i\alpha} = F^{\rm [I]}_{i\alpha} + F^{\rm [II]}_{i\alpha} +F^{\rm
    [III]}_{i\alpha} 
\end{equation}

where the second term is the force due to the radial derivatives of
$v_2$, the first and third components are the force due to the density
derivatives of $E_{\rm vol}$ and $v_2$ respectively. The first
component can be written 

\begin{equation}
  F_{i\alpha}^{\rm [I]} = -\pd{E_{\rm vol}}{\bar{n}_i}
  \pd{\bar{n}_i}{R_{i\alpha}} - \sum_{j \ne i} \pd{E_{\rm
      vol}}{\bar{n}_j} \pd{\bar{n}_j}{R_{i\alpha}} 
  \label{eq:fia1}
\end{equation}

where $\partial E_{\rm vol} / \partial \bar{n}_i$ is shorthand for the
density derivative evaluated at $\bar{n}_i$. We can write the
derivatives of the spatially-averaged local electron density, noting
that the on-site density $\bar{n}_a$ does not contribute, as 

\begin{equation}
  \pd{\bar{n}_i}{R_{i\alpha}} = \sum_{j \ne i} \pd{\bar{n}_{\rm pa}}{R_{ij}}
  \frac{R_{ji\alpha}}{R_{ij}} 
\end{equation}

and

\begin{equation}
  \pd{\bar{n}_j}{R_{i\alpha}} = \pd{\bar{n}_{\rm pa}}{R_{ij}}
  \frac{R_{ji\alpha}}{R_{ij}} 
\end{equation}

where $R_{ji\alpha}$ is the $\alpha$ component of the difference
between position vectors ${\bf R}_i - {\bf
  R_j}$ and $R_{ji\alpha}/R_{ij}$ are the direction
cosines. Eq.\ref{eq:fia1} can be in more symmetric form  

\begin{equation}
  F_{i\alpha}^{\rm [I]} = \sum_{j \ne i} \left( \pd{E_{\rm
      vol}}{\bar{n}_i} + \pd{E_{\rm vol}}{\bar{n}_j} \right)
  \pd{\bar{n}_{\rm pa}}{R_{ij}} \frac{R_{ij\alpha}}{R_{ij}}. 
\end{equation}

The second component of the force looks similar to the GPT
force. However, it is only equal to the GPT force in the bulk. It is
given by 

\begin{equation}
  F^{\rm [II]}_{i\alpha} = \sum_{j \ne i} \pd{v_2}{R_{ij}}
  (R_{ij},\bar{n}_{ij}) \frac{R_{ij\alpha}}{R_{ij}}. 
\end{equation}

The final component, which contains an additional neighbor sum, is
given by 

\begin{align}
  F^{\rm [III]}_{i\alpha} &= \frac{1}{2} \sum_{j \ne i}
  \pd{v_2}{\bar{n}_{ij}} \left( \pd{\bar{n}_i}{R_{i\alpha}} +
  \pd{\bar{n}_j}{R_{i\alpha}} \right) \nonumber \\ 
  & \quad + \frac{1}{4} \sum_{j \ne i} \sum_{k \ne j \ne i}
  \pd{v_2}{\bar{n}_{jk}} \left( \pd{\bar{n}_j}{R_{i\alpha}} +
  \pd{\bar{n}_k}{R_{i\alpha}} \right). 
\end{align}

\begin{figure}[t]            
  \includegraphics[width=0.45\textwidth]{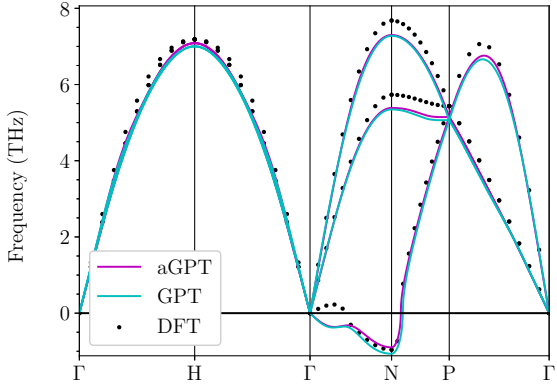}
  \caption{\label{fig:bccph}Phonon dispersion relation for mechanically
    unstable bcc Mg at the equilibrium volume. The averaging sphere
    radius was taken to be $R_a = 1.8 R_{\rm WS}$. The aGPT phonon
    band structure (magenta) is quantitatively similar to the GPT (cyan)
    deviating only in the imaginary sector between high-symmetry
    points $\Gamma$ and $N$. The DFT data (black points) is in good
    agreement with the aGPT/GPT results. The small qualitative
    difference in the imaginary sector along $\Gamma$-$N$ is
    associated with subtle differences in pressure.}   
\end{figure}

\begin{figure}[b]            
  \includegraphics[width=0.45\textwidth]{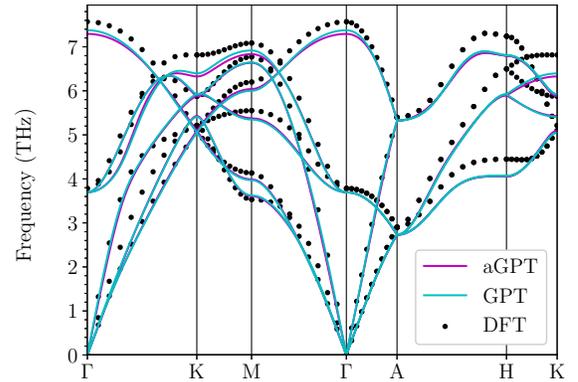}
  \caption{\label{fig:hcpph}Phonon dispersion relation for hcp Mg at
    the equilibrium volume and $c/a$ ratio. The averaging sphere
    radius was taken to be $R_a = 1.8 R_{\rm WS}$. The aGPT phonon
    bandstructure (magenta) is quantitatively similar to the GPT
    (cyan). Both the aGPT and GPT are in good qualitative agreement
    with the DFT data (black dots).}
\end{figure}

The bulk force constant matrix $A_{ij\alpha\beta}$ will largely be the
same as for the GPT. However, there will be small contributions from
the density derivatives of $E_{\rm vol}$ and $v_2$. These additional
contributions require further neighbor summations. These third and
fourth-order terms can be necessary to capture the phonon dispersion at
certain ${\bf q}$-points in the Brillouin zone, in particular for
Be\cite{Bertoni1973}. Despite this, the phonon dispersion will be
dominated by the bulk GPT force constant matrix. However, deviations
in the band structure in the low ${\bf q}$ limit are expected and
correspond to changes in the elastic constants. The phonon dispersion
was calculated for mechanically unstable bcc Mg at the equilibrium
atomic volume in Fig.\ref{fig:bccph}. This crystal structure was
chosen as a representative example due to the presence of the
imaginary frequencies along the ${\bf q}$-point path from $\Gamma$ to
N. There is also scientific interest in this particular phase. When Mg
is alloyed with Li, the bcc phase is stabilised and the alloy becomes
ductile. In addition, the phonon dispersion relation for
thermodynamically stable hcp Mg at the equilibrium atomic volume and
$c/a$ ratio is shown in Fig.\ref{fig:hcpph}. The dispersion relations
were calculated numerically with the code ALAMODE\cite{Tadano2014}
using supercells that were extended by 6x6x6 (for aGPT/GPT) and 3x3x3
(for DFT). The DFT results were calculating using the FP-LMTO method
of van Schilfgaarde and co-workers\cite{Methfessel1999} with the same
lattice parameters as the aGPT/GPT. The Brillouin zone integrations
were performed with Methfessel-Paxton sampling \cite{Methfessel1989}
and 30x30x30 ${\bf q}$-point subdivisions. The exchange-correlation
functional was taken in the local-density approximation using the
correlation function of Perdew and Wang \cite{Perdew1992}. The local
density approximation was used since it is closest to the treatment of
exchange and correlation within the GPT. There is good agreement with
the DFT data and the aGPT/GPT.

\subsection{Stress Tensor}
\label{sec:stress}

Molecular dynamics simulations that sample an isobaric ensemble
require a barostat to match the external pressure to the internal
pressure\cite{Andersen1980,*Hoover1985} $P^{\rm int} = -\sum_\alpha
\sigma_{\alpha \alpha} / d$ where $\sigma$ is the internal stress
tensor and $d$ is the dimension of the cell. This matching is
effectively the equilibrium condition i.e. the time average of the
internal pressure is the external pressure. In such simulations, only
the lattice parameter $a$ is dynamic. This constraint is slightly
artificial if the crystal has multiple lattice parameters as in the
case of hexagonal crystals. Relaxing this constraint requires that we
now sample an isostress ensemble \cite{Parrinello1981} where the
internal stress tensor is matched to an external stress tensor. The
stress tensor is defined as the infinitesimal change in total energy
as a result of an infinitesimal strain\cite{Wallace1998} 

\begin{equation}
  \sigma_{\alpha\beta} = \frac{1}{V} \left( \pd{E_{\rm
      tot}}{\varepsilon_{\alpha\beta}} \right) \Bigg
  |_{\varepsilon_{\alpha\beta} = 0} 
  \label{eq:stress}
\end{equation}

where the prefactor of inverse volume $V^{-1}$ is required by
dimensional analysis. The application of a strain changes the lattice
vectors $h$ in the following way 

\begin{equation}
  h_{\alpha\beta} \to \tilde h_{\alpha \beta} = \sum_{\gamma} \left(
  \delta_{\alpha\gamma} + \varepsilon_{\alpha\gamma} \right)
  h_{\gamma\beta} 
\end{equation}

where $\delta$ is the Kronecker delta. Since the lattice vectors act
as basis vectors for the position vectors of the atoms, a strain
transforms the ion at site $i$ to a new position $\tilde R_{i\alpha}$
i.e.  

\begin{equation}
  R_{i\alpha} \to \tilde R_{i\alpha} = \sum_\beta \tilde h_{\alpha
    \beta} S_{i\beta} = R_{i\alpha} + \sum_{\beta\gamma}
  \varepsilon_{\alpha\beta} h_{\beta \gamma} S_{i\gamma} 
  \label{eq:strij}
\end{equation}

where $S_{i\alpha}$ is the position of site $i$ in a fractional
co-ordinate system. After application of the strain, the Cartesian
distance between sites $i$ and $j$ is given by 

\begin{equation}
  \tilde R_{ij} = \sqrt{\sum_{\alpha\beta} \tilde G_{\alpha\beta}
    (\varepsilon) S_{ij\alpha} S_{ij\beta}} 
\end{equation}

where $\tilde G_{\alpha\beta} (\varepsilon) = \sum_\gamma \tilde
h_{\gamma\alpha} \tilde h_{\gamma\beta}$ is the strained metric
tensor. If the strain is sufficiently small so as to vanish at
quadratic order $\varepsilon_{\alpha\beta} = \delta
\varepsilon_{\alpha\beta}$, we may write 
 
\begin{equation}
  \tilde G_{\alpha\beta} (\varepsilon) = G_{\alpha\beta} + 2
  \sum_{\mu\nu} \delta \varepsilon_{\mu\nu} h_{\mu\alpha} h_{\nu\beta} 
\end{equation}

where $G_{\alpha\beta}$ is the metric tensor of the unstrained
crystal. By denoting the second term as $\delta G_{\alpha\beta}$ and
expanding Eq.\ref{eq:strij} about $\delta G = 0$, we find 

\begin{equation}
  \tilde R_{ij} = R_{ij} + \frac{1}{2 R_{ij}} \sum_{\alpha\beta}
  \delta G_{\alpha\beta} S_{ij\alpha} S_{ij\beta}. 
\end{equation}

This Taylor expansion allows us to explicitly evaluate the derivative
of the interatomic separation $R_{ij}$ with respect to strain 
 
\begin{equation}
  \pd{R_{ij}}{\varepsilon_{\alpha\beta}} = \lim_{\delta
    \varepsilon_{\alpha\beta} \to 0} \left[ \frac{\tilde R_{ij} -
      R_{ij}}{\delta \varepsilon_{\alpha\beta}} \right] =
  \frac{R_{ij\alpha} R_{ij\beta}}{R_{ij}}. 
\end{equation}

Turning to the aGPT stress tensor, we make a decomposition of the
stress tensor $\sigma$ along the same lines as for the force

\begin{equation}
  \sigma_{\alpha\beta} = \sigma^{[\rm I]}_{\alpha\beta} + \sigma^{[\rm
      II]}_{\alpha\beta} + \sigma^{[\rm III]}_{\alpha\beta}. 
\end{equation}

The second term in the decomposition takes the form of a virial stress
tensor 

\begin{equation}
  \sigma^{[\rm II]}_{\alpha\beta} = \frac{1}{2V} \csum_{ij} {\cal
    F}^{\rm [II]}_{ij\alpha} R_{ij\beta}   
\end{equation}

where ${\cal F}^{[\rm II]}_{ij\alpha}$ is defined as the force on ion
$i$ due to ion $j$  

\begin{equation}
  F^{[\rm II]}_{i\alpha} = \sum_{j \ne i} {\cal F}^{\rm [II]}_{ij\alpha}.
\end{equation}

Including volume dependence of the spatially-averaged pseudoatom
density will mean that first contribution to the total stress tensor
cannot be written as a virial

\begin{equation}
  \sigma^{\rm [I]}_{\alpha\beta} = \frac{1}{V} \sum_i \pd{E_{\rm
      vol}}{\bar{n}_i} \pd{\bar{n}_i}{\varepsilon_{\alpha\beta}} 
  \label{eq:str1}
\end{equation}

where, using the identity $\partial \Omega / \partial
\varepsilon_{\alpha\beta} = \Omega \delta_{\alpha\beta}$, we have

\begin{align}
  \frac{1}{V} \pd{\bar{n}_i}{\varepsilon_{\alpha\beta}} &= \frac{1}{N}
  \left( \pd{\bar{n}_a}{\Omega} + \sum_{j \ne i} \pd{\bar{n}_{\rm
      pa}}{\Omega} (R_{ij},\Omega) \right) \delta_{\alpha\beta} \nonumber \\
  & \quad + \frac{1}{V} \sum_{j \ne i} \pd{\bar{n}_{\rm pa}}{R_{ij}}
  \frac{R_{ij\alpha} R_{ij\beta}}{R_{ij}}.
\end{align}

This form can be inserted into Eq.\ref{eq:str1} and made more
explicitly symmetric in $i$ and $j$

\begin{align}
  \sigma_{\alpha\beta}^{\rm [I]} &= \frac{1}{N} \pd{\bar{n}_a}{\Omega}
  \sum_i \pd{E_{\rm vol}}{\bar{n}_i} \delta_{\alpha\beta} \nonumber \\ 
  & \quad + \frac{1}{2N} \csum_{ij} \left( \pd{E_{\rm vol}}{\bar{n}_i}
  + \pd{E_{\rm vol}}{\bar{n}_j} \right) \pd{\bar{n}_{\rm pa}}{\Omega}
  \delta_{\alpha\beta} \nonumber \\
  & \quad \quad \frac{1}{2V} \csum_{ij} \left( \pd{E_{\rm
      vol}}{\bar{n}_i} + \pd{E_{\rm vol}}{\bar{n}_j} \right)
  \pd{\bar{n}_{\rm pa}}{R_{ij}}
  \frac{R_{ij\alpha}R_{ij\beta}}{R_{ij}}. \nonumber
\end{align}

The final contribution to the total stress tensor is given by

\begin{equation}
  \sigma^{\rm [III]}_{\alpha\beta} = \frac{1}{4 V} \sum_i \sum_{j\ne
    i} \pd{v_2}{\bar{n}_{ij}} \left(
  \pd{\bar{n}_i}{\varepsilon_{\alpha\beta}} +
  \pd{\bar{n}_j}{\varepsilon_{\alpha\beta}} \right).
\end{equation}

We calculate the elastic constants numerically by approximating the
derivative  

\begin{equation}
  C_{\alpha\beta\gamma\delta} =
  \left(\pd{\sigma_{\alpha\beta}}{\varepsilon_{\gamma\delta}}\right)
  \Bigg|_{\varepsilon_{\alpha\beta} = 0}. 
  \label{eq:elastic}
\end{equation}

Since the elastic constants are extremely sensitive to minor changes
in the potential, we choose to approximate the derivative in
Eq.\ref{eq:elastic} using a central difference method whose error is
of quartic order in the strain parameter. These results are shown
alongside GPT and experimental values in Table \ref{tab:elastic}. The
dynamical bulk modulus $B_d$ is calculated by using combinations of
volume-dependent elastic constants. The aGPT values $B_d = 35.9
~\rm{GPa}$ in Table \ref{tab:elastic} are in excellent agreement with
the static bulk modulus $B_s = 35.8 ~\rm{GPa}$ that was calculated
from derivatives of the equation of state. \par

The elastic constants can be used to find an optimum value of the
averaging sphere radius $R_a$. In particular, the volume-conserving
elastic constants should be equivalent in the GPT and aGPT. The
difference arises as a result of the approximations and assumptions
made in the aGPT formalism. With reference to our calculated aGPT
elastic constants in Table \ref{tab:elastic}, a smaller cut-off radius
$R_a = 1.8 R_{\rm WS}$ better reproduces the volume-conserving GPT
elastic constants. Also note from Table \ref{tab:elastic} that the
compressibility problem is removed, with both the GPT and aGPT values
of the dynamic bulk modulus $B_d$ in good agreement with the static
value $B_s = 35.8 ~\rm{GPa}$. 

\begin{table}[t]
  \caption{\label{tab:elastic} Elastic constants calculated for hcp
    Mg with the experimentally observed values for $\Omega =
    156.8 ~\rm{a.u.}$ and $c/a = 1.62$. The GPT elastic constants were
    calculated in two ways, using only the virial stress tensor
    without basal plane relaxation (labelled `Virial' in the table) and
    using the virial stress tensor with volume derivatives and basal
    plane relaxation. The aGPT elastic constants were calculated at
    two physically reasonable values of the averaging-sphere
    radius. The experimental values\cite{Slutsky1957} were measured at
    $300 ~{\rm K}$.}
  \begin{ruledtabular}
    \begin{tabular}{ c  c  c  c  c  c  c  c}
      [$\rm{GPa}$]         & $C_{11}$ & $C_{12}$ & $C_{13}$ & $C_{33}$ &
      $C_{44}$ & $C_{66}$ & $B_d$ \\
      \hline
      GPT (Virial)       & 73.2 & 27.8 & 24.6 & 63.6 & 19.5 & 22.7 & 40.5 \\
      GPT                & 63.9 & 25.2 & 21.1 & 62.6 & 19.5 & 19.4 & 36.1 \\
      aGPT ($R_a = 1.8$) & 63.5 & 25.5 & 20.6 & 62.7 & 19.5 & 19.0 & 35.9 \\
      aGPT ($R_a = 3.4$) & 63.3 & 25.3 & 21.0 & 61.6 & 18.9 & 19.0 & 35.9 \\
      Exp.\cite{Slutsky1957} & 59.4 & 25.6 & 21.4 & 61.6 & 16.4 & 16.9 & 35.2
    \end{tabular}
  \end{ruledtabular}
\end{table}

\section{Results}
\label{sec:results}

\subsection{Vacancy Formation Energy}
\label{sec:vfe}

\begin{table}[t]
  \caption{\label{tab:mono} Vacancy formation energies calculated for
    hcp Mg. The vacancy formation energy $E^{1f}$was calculated using
    both the GPT and the aGPT for two values of the averaging sphere
    radius at the GPT equilibrium atomic volume $\Omega = 156.8
    ~\rm{a.u.}$ and $c/a = 1.62$. The DFT values \cite{Uesugi2003}
    were calculated at the zero temperature lattice parameters
    (excluding zero point phonons).}
  \begin{ruledtabular}
    \begin{tabular}{ c  c  c  c  c  c  c}
      [eV]         & $v_2^{1\rm f}$ & $E_{\rm vol}^{1\rm f}$ & $\Delta
      v_2^{1\rm f}$ & $E_{\rm rlx}^{1\rm f}$ & $E^{1\rm{f}}$ &
      $\Omega^{1\rm f}$ \\
      \hline
      GPT & 0.44 & 0.00 & - & -0.01 & 0.43 & 0.71 \\
      aGPT ($R_a = 1.8$) & 0.44 & 0.47 & -0.19 & -0.01 & 0.71 & 0.65 \\
      aGPT ($R_a = 3.4$) & 0.44 & 0.50 & -0.23 & -0.01 & 0.70 & 0.59 \\
      DFT\cite{Uesugi2003} & - & - & - & -0.01 & 0.74 & 0.69 \\
      Exp. & - & - & - & - & $0.79\pm0.03$ \cite{Tzanetakis1976} & - \\
    \end{tabular}
  \end{ruledtabular}
\end{table} 

A vacancy is the primary test case for the aGPT, since it is the
simplest defect for which there is considerable local volume
change. As a consequence, a large amount of the energy required
to create a vacancy is not captured by the GPT and other methods based
upon second-order pseudopotential perturbation theory. The vacancy
formation energy is usually defined as the energy required to remove
one atom to infinity and replace it at the surface. The vacancy
formation energy $E^{1\rm f}$ can be written without
approximation\cite{Finnis2003} as

\begin{equation}
  E^{1\rm f} = \lim_{N_a \to \infty} \left[ E_{\rm tot} (N_a,1) -
    \left(\frac{N_a-1}{N_a}\right) E_{\rm tot} (N_a,0) \right]
  \label{eq:mono}
\end{equation} 

where $N_a$ is the number of sites and $E_{\rm tot}$ is a function of 
both the number of atoms and number of vacancies. The term in the
brackets can be evaluated at finite $N$ and then extrapolated into the
thermodynamic limit $N \to \infty$. Provided that the atomic positions
are relaxed and we are using the bulk equilibrium lattice parameters,
it is unnecessary to relax the lattice parameters for the vacancy
cell. This is because the largest correction to the vacancy formation
energy is $-P \Omega^{1 \rm f}$ where $\Omega^{1 \rm f}$ is the misfit
or vacancy formation volume. \par

We have calculated the relaxed vacancy formation energy in hcp Mg at
the experimentally observed atomic volume $\Omega = 156.8 ~{\rm a.u.}$
and $c/a = 1.62$. In our calculations, the atomic volume is kept
constant which means that the removal of an atom gives rise to a
contraction of the lattice. The vacancy formation energy is calculated
at multiple values of $N$ and extrapolated to the thermodynamic
limit. In addition, we also calculate the misfit volume $\Omega^{1 \rm
  f}$ using the following formula \cite{Moriarty1991}

\begin{equation}
  \Omega^{1 \rm f} / \Omega_0 = -B_d^{-1} \pd{E^{1\rm f}}{\Omega}
\end{equation} 

where $B$ is the bulk modulus as calculated in Section
\ref{sec:derivs}. These results are given in Table \ref{tab:mono} and
compared to GPT and experimental vacancy formation energies. The
vacancy formation energy was calculated for $N_a \in \{54, 128, 250,
432\}$ and then extrapolated to infinity. The extrapolated vacancy
formation energy is around 1\% less than the vacancy formation energy
for $N_a = 432$. The divacancy binding energy was also calculated for
hcp Mg using the following formula

\begin{equation}
  E^{2\rm b}_{i \rm{NN}} = 2 E^{1\rm f} - E^{2\rm f}_{i \rm{NN}}
\end{equation}

where $E^{2\rm f}_{i \rm{NN}}$ is the divacancy formation energy for a
vacancy at the origin and a vacancy in the $i^{\rm th}$ neighbor
shell. The divacancy formation energy was calculated using an
analogous expression to Eq.\ref{eq:mono}. The ordering of the first
and second nearest neighbors is dependent on the $c/a$ ratio in hcp
crystals. In Mg, the $c/a$ ratio is less than the ideal value which
means that the first nearest neighbor lies at a distance less than
the lattice parameter $a$. We make a nearest neighbor definition
along the same lines as Uesugi {\it et al.}\cite{Uesugi2003} The
divacancy binding energy compiled in Table \ref{tab:di}, converges
more slowly with $N_a$ than the vacancy formation energy. In addition,
the divacancy binding energy converges more slowly for the aGPT than
it does for the GPT. As such, the divacancy binding energy was
calculated for larger values of $N_a \in \{250,432,686,1024\}$. Both
the aGPT and the GPT are under bound over the first two neighbor
shells relative to DFT. We note however, that the divacancy binding
energy is the difference between two quantities with unknown error
bars. Therefore, it is unclear whether the underbinding of the aGPT is
a deficiency of the method. \par 

\begin{table}[b]
  \caption{\label{tab:di} Relaxed divacancy binding energies
    calculated for hcp Mg with the equilibrium GPT values for $\Omega$
    and $c/a$. We have calculated $E^{2b}_{i\rm NN}$ using the GPT and
    the aGPT for two values of the averaging sphere radius. The DFT
    values \cite{Uesugi2003} were calculated at the zero temperature
    lattice parameters (excluding zero point phonons).}
  \begin{ruledtabular}
    \begin{tabular}{ c  c  c  c  c  c}
      [eV] & $E_{1\rm NN}^{2b}$ & $E_{2\rm NN}^{2b}$ & $E_{3\rm
        NN}^{2b} $ & $E_{4 \rm NN}^{2b}$ & $E_{5 \rm NN}^{2b}$ \\
      \hline
      GPT & $+0.02$ & $+0.02$ & $-0.01$ & $+0.00$ & $+0.00$ \\
      aGPT ($R_a = 1.8$) & $+0.01$ & $+0.01$ & $-0.05$ & $-0.03$ & $-0.02$ \\
      aGPT ($R_a = 3.4$) & $+0.00$ & $+0.01$ & $-0.02$ & $-0.01$ & $-0.01$ \\
      DFT\cite{Uesugi2003} & $+0.06$ & $+0.07$ & $-0.01$ & $+0.01$ & $+0.01$
    \end{tabular}
  \end{ruledtabular}
\end{table}

\subsection{Stacking Fault Energies}
\label{sec:sfe}

\begin{figure}[t]
  \includegraphics[width=0.45\textwidth]{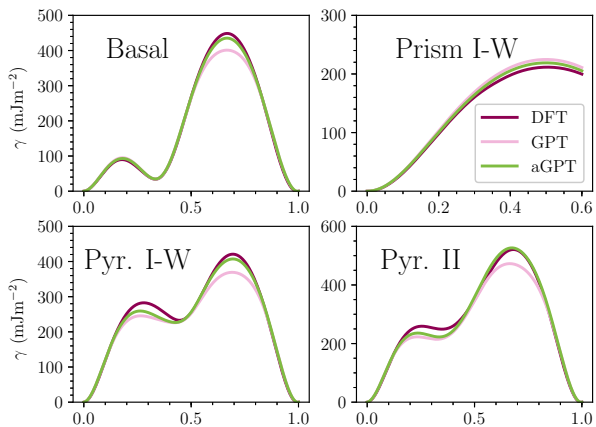}
  \caption{\label{fig:sfe}The $\gamma$-line calculated for hcp Mg
    using the GPT (pink), aGPT (green) and compared to the DFT results
    (red) of Yin \textit{et al} \cite{Yin2017}. For comparison, the
    GPT and aGPT was calculated using the DFT zero temperature lattice
    parameters. The crystal was tilted along the $[1\bar{1}00]$,
    $1/3[\bar{1}2\bar{1}0]$, $1/2[\bar{1}102]$ and $1/3[\bar{2}113]$
    in the Basal, Prism I, Pyramidal I and II crystallographic planes
    respectively. The aGPT and GPT are in agreement for the stable
    stacking fault energy. The aGPT increases the value of the
    unstable stacking fault relative to the GPT.}
\end{figure}

Information about the plastic behavior of a metal can be inferred from a
calculation of the stacking fault energies and the profile of the
$\gamma$-line. In particular, the stacking fault energy controls the
dissociation width of dislocations into partial dislocations. This in
turn controls the ability of a dislocation to cross-slip and limits
easy-glide. The $\gamma$-line is defined in the following manner. An
infinite crystal is partitioned into two subcrystals with their
interface being some crystallographic plane. One half of the crystal
is moved relative to the other along some crystallographic direction
until the crystal has been translated by an integer multiple of the
lattice vectors. The $\gamma$-line is the relative energy change
during this process, normalized by the area of the crystallographic
plane. \par

Practical computations pose several challenges for this
procedure. All of the approaches begin by choosing a supercell whose
lattice vectors $\{{\bf a}_1, {\bf a}_2\}$ define the crystallographic
plane over which the slip occurs. For instance, in the basal plane of
the hcp structure these can be represented as the Cartesian vectors
${\bf a_1} = [1,0,0]$ and ${\bf a_2} = [-1/2,\sqrt{3},0]$. The
supercell is extended $n$ times in the ${\bf a}_3$ direction such that
there are $n$ unit cells . The definition of ${\bf a}_3$ is not unique
and it need not be perpendicular to the crystallographic plane. In
fact, the only requirement on ${\bf a}_3$ is that it connects to an
atom which is out of the crystallographic plane. There are a number of
ways to create the stacking fault. One such method is the so-called
`slab' method\cite{Vitek1968} whereby the stacking fault is created by
moving atoms relative to each other at the approximate center of the
supercell. With periodic boundary conditions, the `slab' method
creates an additional stacking fault at the boundary of the supercell
with the periodic image. Another method, which we employ, creates the
fault by tilting the out-of-plane lattice vector ${\bf a}_3 \to {\bf
  a}_3 + \alpha {\bf t}$ where ${\bf t}$ is some integer combination
of the in-plane lattice vectors and $\alpha$ is a real number in the
interval $[0,1]$. The `tilt' method creates only one stacking fault
per supercell whereas the `slab' method creates two. Thus, with the
tilt method there is faster convergence with the number of unit cells
$n$. \par 

If a crystal has a stacking fault, the atoms will relax in order to
minimize the interatomic forces that were created by the
fault. Using the original V{\'{\i}}tek description\cite{Vitek1968} of
the $\gamma$-line, only out-of-plane relaxations are allowed. If such
restrictions were not in place then the atoms would relax to either
the equilibrium positions or the stable stacking fault up to some
strain due to the finite supercell. In certain crystallographic
planes and for certain elements, notably the Pyramidal II plane for
Mg\cite{Yin2017}, both the stable stacking fault energy and stacking
fault vector calculated using the V{\'{\i}}tek method are not very
close to the fully relaxed values. Along these planes if the entire
$\gamma$-line is desired then it is necessary to remove the
restrictions on in-plane relaxations away from the fault
itself\cite{Morris1997} or using a nudged elastic band method. The
aGPT $\gamma$-line was calculated using the V{\'{\i}}tek method for
hcp Mg along 4 directions in 4 crystallographic planes in
Fig.\ref{fig:sfe} for both the GPT and aGPT. In general, we find that
there is agreement between the GPT and aGPT at the stable stacking
fault. However, for the unstable stacking fault the aGPT improves upon
the GPT relative to the DFT results of Yin {\it et al.}\cite{Yin2017}

\subsection{Finite Temperature Lattice Parameters}
\label{sec:ftlp}

\begin{figure}[t]
  \includegraphics[width=0.45\textwidth]{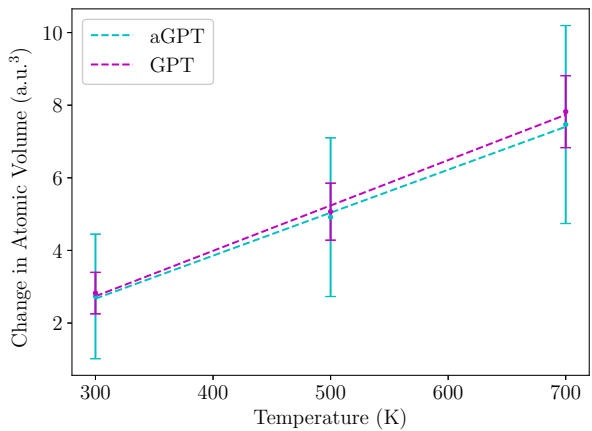}
  \caption{\label{fig:latp}The volume for hcp Mg was averaged over an
    $N$$P$$T$ ensemble was calculated using molecular dynamics for the
    GPT and aGPT at three temperatures. We have plotted the change in
    atomic volume relative to the zero temperature value. The error
    bars for the 95\% confidence interval were found using a bias-free
    estimator of the variance. We have plotted the change in atomic
    volume relative to the zero temperature value.}
\end{figure}

Whilst we expect the aGPT to apply well to finite temperature, thanks
to good agreement with the GPT and DFT harmonic phonon band structure
in Figs. \ref{fig:bccph} and \ref{fig:hcpph}, it is important to
assess its ability to describe anharmonic effects too. We have looked
at thermal expansion, since it is not well captured by quasiharmonic
lattice dynamics. For instance, Althoff \textit{et al.}
\cite{Althoff1993} calculated the thermal expansion coefficient
$\beta$ in the quasiharmonic approximation with the GPT and found that
there was a discrepancy of roughly 33\% between the quasiharmonic
values and experimental values. However, close agreement to experiment
was found when anharmonic corrections were added in. In this work, we
have calculated the volume in hcp Magnesium at finite temperature with
fixed $c/a$ ratio using molecular dynamics and the stochastic
thermostat and barostat of Bussi, Zykova-Timan and Parrinello
\cite{Bussi2009}. For both the GPT and aGPT, we ran 8 simulations with
different initial velocities corresponding to separate draws from the
Maxwell-Boltzmann distribution at 3 temperatures and zero
pressure. These simulations ran for 40000 time steps, with a further
40000 time steps for equilibration, and a time step of $0.1
~\mathrm{fs}$. The aGPT calculation was performed with 512 atoms
whilst the GPT calculation was performed with 2000 atoms. The results
of this calculation are plotted in Fig.\ref{fig:latp}. In addition, we
estimate the thermal expansion coefficient by regressing the volume on
the temperature. For the GPT at 500 $\mathrm{K}$, we calculate $\beta
= 7.56 \times 10^{-5} ~\mathrm{K}^{-1}$ and for the aGPT, we calculate
$\beta = 7.17 \times 10^{-5} ~\mathrm{K}^{-1}$. These results are in
excellent agreement with the previous results of Althoff \textit{et
  al}. \cite{Althoff1993} and experimental values \cite{Rao1974}.   
  
\section{Conclusions}
\label{sec:conc}

We have demonstrated that it is possible to include local volume
effects by modifying the GPT so that it now depends on a
spatially-averaged local electron density. In particular, we have
developed the aGPT formalism to the extent that it is now possible to
do molecular statics and dynamics. To this end, we calculated the
vacancy formation energy in hcp Mg at the equilibrium lattice
parameters. The aGPT relaxed vacancy formation energy significantly
improves upon the relaxed GPT vacancy formation energy relative to the
experimentally observed value. In addition, the aGPT provides improved
stacking fault energies for hcp Mg.\par

The computational cost in time of the aGPT is greater than that of the
GPT. This is a result of the additional neighbor loop in the
calculation of the forces. Provided that the neighbor table maker is
linear scaling ${\cal O} (N)$, for instance using a linked list, both
the GPT and aGPT are ${\cal O} (N)$. Relative to empirical potentials,
the major computational cost is due the long range cut-off in both the
GPT and aGPT. This can be demonstrated by considering a short ranged
empirical pair potential whose neighbor cut-off is roughly $1/3$ that
of the GPT (i.e. it runs over the first handful of neighbor shells),
we would expect the GPT to be approximately $3^3 = 27$ times
slower. Furthermore, we expect the aGPT to be $27 N_c$ slower than the
GPT where $N_c$ is the number of atoms in a linked-list block.   
\par

Bulk properties such as phonon dispersion and elastic constants were
also calculated as fundamental tests of the aGPT. The inclusion of the
spatially-averaged local electron density modifies the bulk phonon
dispersion. This is a result of the additional derivatives of the
electron density that appear in the expression for the force constant
matrix. The elastic constants are also modified although the
volume-conserving elastic constants should be the same as for the
GPT. It is only the assumptions and approximations in the aGPT that
make them differ. Thus, we can use the volume conserving elastic
constants to find an optimum value for the averaging sphere radius
$R_a$ which is the lone free parameter in the aGPT. This constraint
would appear to favor near-neighbor values of $R_a$, for instance
$R_a = 1.8 R_{\rm WS}$. \par 

The aGPT can be used to accurately calculate self-diffusion and
defect-defect interactions in elemental metals. However, further work
needs to be done on extending the aGPT to alloys in order to study
solute diffusion or solute-defect interactions. We plan to use the
aGPT to further study vacancies and, in particular, the
high-temperature deviation from Arrhenius behavior
\cite{Glensk2014}. All of the results presented in this paper were
calculated using our in-house Fortran codes. There is a planned future
project to incorporate the aGPT into LAMMPS \cite{Plimpton1995}.

\begin{acknowledgments}
  The authors would like to thank Prof. Mike Finnis for thought
  provoking discussions, Prof. Bill Curtin for providing comments on
  an early version of the manuscript and Dr Zhaoxuan Wu for providing
  the DFT stacking fault data. G.C.G.S. acknowledges support from the
  UK EPSRC under the Doctoral Training Partnership
  (DTP). A.T.P. acknowledges the support of the UK EPSRC under the
  grant Designing Alloys for Resource Efficiency (DARE),
  EP/L025213/1. The work of J.A.M. was performed under the auspices of
  the U.S. Department of Energy by Lawrence Livermore National
  Laboratory under Contract No. DE-AC52-07NA27344.   
\end{acknowledgments}

%\appendix*

\bibliographystyle{apsrev4-1}
\bibliography{biblio}

\end{document}